\documentclass[twocolumn,doublespacing,showpacs,preprintnumbers,nofootinbib,amsmath,amssymb]{revtex4}
\usepackage{mathtext}
\usepackage{graphics}
\usepackage{epsfig,amsfonts,amssymb}% Include figure files
\usepackage{bm}% bold math
\usepackage{multirow}

\newcommand\Lam{\Lambda}

\newcommand\half{{\frac{1}{2}}}

\newcommand\lam{\lambda}

\begin{document}

%\title{Near-threshold pion production and $\bm{P}$-wave dibaryon resonances}% Force line breaks with \\
\title{ $\bm{P}$-wave dibaryon resonances in $\bm{pp}$ elastic scattering \\ and near-threshold pion production}
\author{O.A. Rubtsova }%
\email{rubtsova-olga@yandex.ru}
\author{V.I. Kukulin}
\email{kukulin@nucl-th.sinp.msu.ru}
\author{M.N. Platonova }%
\email{platonova@nucl-th.sinp.msu.ru}
\affiliation{%%
Skobeltsyn Institute of Nuclear Physics, Lomonosov Moscow State
University, Leninskie Gory 1/2, 119991 Moscow, Russia}

\date{\today}% It is always \today, today, %  but any date may be explicitly specified

\begin{abstract}
It is demonstrated within the dibaryon-induced model for $NN$ interaction that $pp$ scattering in $P$ waves is governed mainly by the production of the intermediate dibaryon resonances. Two dibaryon resonances with a mass of about 2200 MeV discovered recently by the
ANKE-COSY Collaboration are shown to determine
both elastic $pp$ phase shifts and inelasticities in the $^3P_0$ and $^3P_2$--$^3F_2$ channels from zero energy up to $T_p=0.7$--$0.9$ GeV. It is also demonstrated clearly that the $^3P_0$ dibaryon plays a
decisive role in near-threshold neutral pion production in $pp$ collisions which is
poorly understood to date. The missing dibaryon
contribution is found to be the very possible reason for the
failure of traditional approaches to explain near-threshold
$\pi^0$ production.
\end{abstract}

%\pacs{03.65.Nk,21.45.-v,25.45.De}
%\begin{keyword}
 %\end{keyword}
%\end{frontmatter}
\maketitle
\section{Introduction}
Meson production in $NN$ collisions is a
long-standing problem in nuclear physics because such processes
include rather high momentum transfers and thus the production
mechanism is tightly interrelated to short-range $NN$ dynamics
which is poorly understood to date. One of the most challenging processes in this area has been the neutral pion production in $pp$ collisions near threshold. Thus, in the early attempts to describe
this process, a surprising
result was attained, i.e., the theoretical cross section was found
to underestimate the respective experimental data by about 5 times
\cite{Koltun-Reitan,Sauer,Meyer}.
%In the following years, this strong
%disagreement has been reduced somehow to about 3 times
%underestimation \cite{EFT}.

On the other hand, in the 1990ies and later, rather accurate and
complete data on the near-threshold total cross section
\cite{Meyer,Bondar,WASA_M} as well as the polarisation observables \cite{Meyer2} in the
$pp\to pp\pi^0$ reaction appeared.
In the experiments \cite{Meyer2}, scattering of a spin-polarised beam off a spin-polarised target was measured. The
new high-precision data stimulated numerous calculations in this area (see reviews~\cite{Hanhart,Baru}).
The calculations within the phenomenological models (see, e.g.,~\cite{Lee,Horowitz,Andreev,Julich,Elster,Oset}) supposed various model-dependent explanations for the observed discrepancies, such as heavy meson exchanges or off-shell corrections to the $\pi N$ amplitude, and were therefore inconclusive. Then the substantial progress in treating pion production was achieved within the Chiral Perturbation Theory (ChPT) \cite{Park,Hanhart2,Filin} indicating that some sizable contributions to the $pp\to pp\pi^0$ cross section can come from the next-to-next-to-leading order terms of the chiral perturbation series. However, a quantitative description of neutral pion production has not been obtained to date.

\begin{figure}[ht]
\centering \epsfig{file=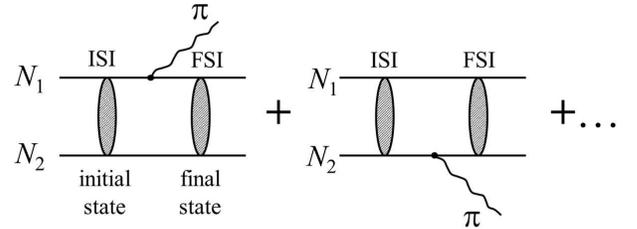,width=\columnwidth}
\caption{\label{fig1} The direct one-nucleon mechanism for near-threshold
pion production in $NN$ collisions. }
\end{figure}

It seems that the true reason for the above discrepancies was that
the conventional mechanisms used to describe the data did not include some important ingredients.
Thus, in Fig.~\ref{fig1} the direct production mechanism is shown. The pion rescattering term is small here, because, contrary to the charged pion production, the intermediate $\Delta$ contribution is suppressed due to conservation laws. In fact, for the diagrams shown in Fig.~\ref{fig1}, there is a strong mismatch in the $NN$
relative momenta in the initial and final states, so that, the pion production operator should
include the high-momentum components. This means high sensitivity of
the respective matrix elements to the $\pi NN$ vertex form factor
and especially to the short-range cutoff parameter $\Lam_{\pi NN}$.
It is known that in
the conventional meson-exchange approaches to the $NN$ interaction
potential, the value of $\Lam_{\pi NN}$ is usually taken
rather high, i.e., $\Lam_{\pi NN}\simeq$ 1--1.7 GeV/$c$ (like in the
Bonn family of $NN$ potentials \cite{Bonn}), and the same high values of this parameter are
used for the meson-exchange currents as well as for the $NN$ wavefunctions
employed for the initial state interaction (ISI) and
final state interaction (FSI) treatment. Contrary to this, all
microscopic meson-nucleon dynamics including $\pi N \to \Delta \to
\pi N$ on-mass-shell scattering employs much lower values of the
cutoff parameters, i.e., $\Lam_{\pi NN}\simeq$ 0.5--0.9
GeV/$c$ and $\Lam_{\pi N\Delta}\simeq$ 0.3--0.4
GeV/$c$~\cite{Koepf,Gari}. The soft cutoff values are also consistent with the lattice-QCD calculations~\cite{Liu}.
It is evident that an artificial increase of these parameters
by 2--3 times leads immediately to the strong enhancement
of the pion production cross sections. We should emphasize in this
regard that just the high cutoff values probably explain
the success of some phenomenological models in describing
near-threshold neutral pion production. E.g., the authors of Ref. \cite{Lee}
achieved good agreement with experimental data on the
near-threshold $pp \to pp \pi^0$ cross section by introducing the
short-range axial exchange charge operator with
$\Lam_{\pi NN} = $1.3 GeV/$c$ taken from the Bonn $NN$
potential.

However, when the physically justified (soft) values for the cutoff parameters are chosen, the proposed short-range contributions get considerably reduced and the meson production cross sections, in particular, in the near-threshold region, turn out to be strongly underestimated.
Moreover, there are conceptual questions about the heavy meson exchange contributions (analogous to the contact terms in ChPT) which should dominate the neutral pion production near threshold as proposed in, e.g., Refs. \cite{Lee,Horowitz,Elster}. From the modern viewpoint, $t$-channel exchange of a meson heavier than the pion between the isolated nucleons at the distances less than the nucleon size is a pure phenomenology not relevant to the real physical picture \cite{Barnes}. The most doubtful point seems the concept of scalar meson exchange, since the lightest scalar ($\sigma$) meson is a very broad resonance which cannot be effectively exchanged between the nucleons. The detailed discussion of these conceptual issues can be found in, e.g.,~\cite{Yaf2019}.

 So, we propose here an alternative short-range mechanism of the $NN$ interaction which is also relevant for pion production. In view of the above arguments, it seems natural to take into account the formation of the intermediate dibaryon resonances in $pp$ collisions, which effectively increase the meson production cross sections due to the long lifetime of the resonances. The relatively long-lived dibaryon states are produced largely due to the effect of hidden color suggested by Brodsky et al. in 1980ies~\cite{Brod} which prevents the resonance decay into hadronic channels. In our case, the effect of hidden color is evident when the $4q$--$2q$ quark-cluster model for dibaryons is used (see Sect. V of the present paper). It is important for the whole our approach that a number of dibaryon resonances in $NN$ system have been discovered to date (see the recent review~\cite{hcl}).

A good illustration to these points can be found in the charged pion production process $pp\to d\pi^+$ at intermediate energies.
In the conventional approaches, this process is dominated by the pion rescattering term with the $\Delta$-isobar excitation in the intermediate state, while the direct production mechanism analogous to that shown in Fig.~\ref{fig1} gives only a small background. However the intermediate $\Delta$ excitation turns out to be strongly
sensitive to the short-range cutoff parameters $\Lambda_{\pi NN}$
and especially $\Lambda_{\pi N\Delta}$. As we have shown in Refs.
\cite{NPA2016,PRD2016}, the above conventional mechanisms with the
soft cutoff parameters consistent with $\pi N$ elastic scattering
give only a half cross section of the $pp\to d\pi^+$ reaction in
two dominating partial waves\footnote{We use here the standard notation ${}^{2S+1}L_Jl$ for the partial-wave channels of the $pp \to d \pi^+$ reaction, where $S$, $L$ and $J$ denote the spin, orbital and total angular momenta of the initial $NN$ system, while $l$ denotes the orbital angular momentum of the final pion.} ${}^1D_2p$ and ${}^3F_3d$ at $T_p =
400$--$800$ MeV. Nevertheless, it was demonstrated \cite{NPA2016,PRD2016} that one can still employ quite
moderate meson-baryon cutoffs for the accurate description of this process
at intermediate energies, however a
non-conventional short-range mechanism for pion production should
be introduced. In this novel mechanism, the basic $NN$ interaction
at short distances is driven by generation of intermediate
dibaryon resonances in respective partial waves
\cite{JPhys2001,KuInt} and thus pions are emitted from the decay
of the intermediate dibaryon resonance state which drives the
interaction in the given $NN$ partial-wave channel.
Thus, when the excitation of the known isovector dibaryon resonances\footnote{Here and below we denote the
dibaryon states according to the $NN$ channel $^{2S+1}L_J$ to which
it may couple.} ${}^1D_2(2150)$
and ${}^3F_3(2220)$ near the $N\Delta$ threshold (in $S$ and $P$
waves, respectively) is added coherently to the $t$-channel
$\Delta$ excitation, we achieve a very good description of the partial cross sections in the respective channels \cite{NPA2016,PRD2016}.

\begin{figure}[ht]
\centering \epsfig{file=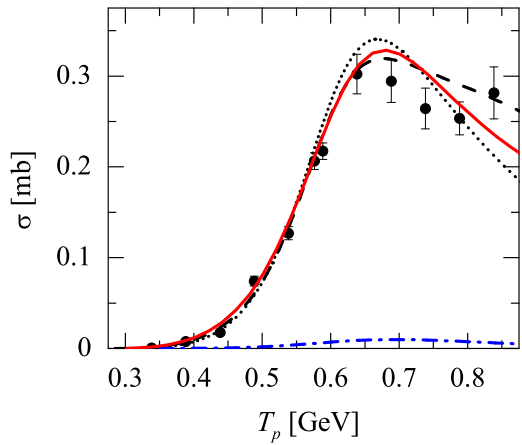,width=0.8\columnwidth}
\caption{\label{fig-3p2d} Partial cross section of the reaction
$pp \to d \pi^+$ in the ${}^3P_2d$ channel. Shown are the summed
contribution of two conventional mechanisms --- pion rescattering with the intermediate
$\Delta$ excitation and direct one-nucleon exchange (dash-dotted curve),
the full calculation including an intermediate dibaryon formation
with parameters $M(^3P_2)=2211$ MeV, $\Gamma(^3P_2)=195$ MeV
(solid curve) and the SAID PWA solutions~\cite{Oh97,SAID}: C500
(dashed curve), SP96 (dotted curve) and single-energy data (filled
circles).}
\end{figure}

The situation is even more crucial for $P$-wave $pp$
scattering with the pion emitted in $d$ wave (the ${}^3P_2d$
partial channel). Here we have an analogy with the near-threshold $\pi^0$ production, where the contributions of both direct and rescattering mechanisms are strongly suppressed. In this channel, the conventional mechanisms give only a
minor (ca. 2.5\%) contribution which is only moderately dependent
on the cutoff parameters. On the other hand, the accurate
description of just the ${}^3P_2d$ amplitude is crucial for
reproducing the experimental data on the polarisation observables
like $A_{xx}$, $A_{yy}$, etc., in $pp \to d \pi^+$ reaction.
Hence, the contribution of the $P$-wave dibaryon ${}^3P_2(2200)$
(which is located also near the $N\Delta$ threshold in the
relative $P$-wave) appears to be {\em the dominant one} in the
${}^3P_2d$ partial channel and is clearly necessary for
reproducing the polarisation observables near the resonance peak
energies $T_p \simeq 600$ MeV. In Fig.~\ref{fig-3p2d} the partial
cross section in this channel is shown in comparison with the partial-wave analysis (PWA) of the SAID group~\cite{Oh97,SAID}.
It is very important that the
${}^3P_2$ dibaryon parameters found from the analysis of the $pp
\to d \pi^+$ reaction turned out to be very close to those found
in a recent experiment of the ANKE-COSY Collaboration~\cite{Komarov16} on a similar reaction $pp
\to \{pp\}_s \pi^0$, where $\{pp\}_s$ is a near-threshold diproton
in the ${}^1S_0$ state.

A quite similar situation takes
place also for the isoscalar double-pion production reactions $pn \to d
(\pi\pi)_0$ \cite{NPA2016,PRC2013}. For now, this is the most bright example of meson production
through excitation of intermediate dibaryons, namely, the $d^*(2380)$ resonance discovered in a
series of experiments of the WASA-at-COSY
Collaboration~\cite{hcl}. While the energies for the $d^*$
excitation in $pn$ collisions $T_p \simeq 1.1$--$1.2$ GeV are
higher than the double-pion production threshold, some interesting
near-threshold phenomenon known as the ABC effect~\cite{ABC} also
takes place here in a sense that the ABC enhancement is located
near the threshold of the dipion invariant-mass
spectrum~\cite{hcl}. In Ref.~\cite{PRC2013} we interpreted this
enhancement as being due to the scalar sigma-meson production with
the mass close to $2m_{\pi}$, which is emitted from the $d^*$
dibaryon and than decays into two final pions. This interpretation
finds support in the measured isospin dependence of the isoscalar
dipion production in $pn$ collisions~\cite{Adl-iso,PK-iso}. On the
other hand, it is generally accepted that the near-threshold
dipion production in $NN$ collisions at energies $T_N < 1$ GeV
goes mainly through the excitation of the intermediate Roper
resonance $N^*(1440)$~\cite{Alvarez}. However it was shown
recently~\cite{Kukulin2,Clement-R} that the formation of the dibaryon
resonance located near the $NN^*(1440)$ excitation threshold
$\sqrt{s} \simeq 2300$ MeV (corresponding to $T_N \simeq 900$ MeV)
is the most likely mechanism for both single- and double-pion
production in the $S$-wave $NN$ collisions. Following these lines,
one could suggest the dibaryon resonances located near (almost)
all nucleonic resonance thresholds to be responsible for the large
portion of the meson production processes which are poorly
reproduced by conventional theoretical models, especially near the
respective production thresholds. For instance, the puzzling
near-threshold $\eta$-meson production in $pp$ and $pd$
collisions(see, e.g.,~\cite{eta}) might be explained by the
intermediate dibaryon excitation near the threshold of
$NN^*(1535)$. Some experimental indications of this dibaryon have
already been obtained~\cite{Ishikawa,Komarov-2}.

It has been shown further in Refs.~\cite{Yaf2019,Kukulin2,FBS2019,Kukulin1} that
the dibaryon generation mechanism supplemented by the peripheral
one-pion exchange gives an accurate description of $NN$ phase shifts and inelasticities in various partial
channels in a broad energy range from zero to about 1 GeV (lab.) and
also leads to the dibaryon resonance states with parameters
very close to their experimental values. The most prominent effect
is seen in the $^3D_3$--$^3G_3$ coupled channels where the
decisive role of the known $d^*(2380)$ dibaryon has been shown at
all energies until $T_N \simeq 1.25$ GeV \cite{Kukulin1}.

In the present study we consider $NN$ scattering in the $^3P_0$, $^3P_1$ and
$^3P_2$--$^3F_2$ partial-wave channels. In two of these channels, viz., $^3P_0$ and $^3P_2$,
the diproton resonances with a mass of about 2200
MeV have been found in the recent experiment of the ANKE-COSY Collaboration
\cite{Komarov16} (the $^3P_2$ resonance was also predicted in the PWA~\cite{Oh97,Arndt}). Here we study the impact of these $P$-wave dibaryons on both elastic and inelastic $pp$ scattering, with a special emphasis on the near-threshold neutral pion production which is governed mainly by the ${}^3P_0$ $pp$ initial state.
%Below we will show within the dibaryon model that just the ${}^3P_0$ dibaryon resonance determines
%the $pp$ elastic phase shift and inelasticity in the ${}^3P_0$ channel at intermediate energies and also governs neutral pion production in the
%near-threshold region.

We especially emphasize here that {\em we do not introduce the dibaryon
resonances ad hoc} to describe the particular meson
production processes but instead we use the recent experimental data \cite{Komarov16} on the $P$-wave dibaryons and the model developed
in the earlier work \cite{Yaf2019,Kukulin2,FBS2019,Kukulin1} which describes
$NN$ scattering at intermediate energies to analyse near-threshold
pion production in $pp$ collisions. Thus, our main goal is to reveal a connection between the $NN$ interaction and fundamental QCD, and neutral pion production near threshold may be considered as an illustration of our general approach.

It should be also noted that the near-threshold charged pion
production with the isoscalar $np$ pair (or the deuteron) in the final state goes mainly via the ${}^3P_1$
$pp$ collisions. However,
the experimental situation with the ${}^3P_1$ dibaryon is not quite clear for now. In particular, it is not
seen as a clear resonance in the respective partial channel of $pp$
elastic scattering or the $pp \to d \pi^+$ reaction and cannot be excited at all in the $pp \to
\{pp\}_0 \pi^0$ process due to the angular momentum and parity
conservation~\cite{Komarov16}. However, some indications of its existence can be found in the literature (see, e.g., Refs.~\cite{Strak3P1,Strak91,MacGregor,Ferreira}). Thus, for consistency of our study of $P$-wave $pp$ scattering within the dibaryon model, we present below the results of calculations for the ${}^3P_1$ partial channel as well.

The structure of the paper is as follows. In Sect. II, we
outline the basic formalism of the dibaryon-induced model for $NN$ interaction and its extension for treatment of inelastic processes. In Sect. III, we present the results of the calculations for the $pp$ elastic scattering phase shifts and inelasticities in the $P$-wave partial channels. In Sect. IV, we discuss in detail the near-threshold neutral pion production within the framework of the
dibaryon model. Sect. V is dedicated to
the feasible microscopic structure of the $P$-wave dibaryon
resonances. Sect. VI summarizes the basic results of the present
work.

\section{Dibaryon-induced model for $\bm{NN}$ interaction}

\subsection{Basic formalism of the model}

In the dibaryon model \cite{Yaf2019,Kukulin2,FBS2019,Kukulin1},
the total Hilbert space includes two channels: an external channel
corresponding to the relative motion of two nucleons and an
additional internal channel which describes the formation of the
six-quark (or dibaryon) state. In the simplest case, the internal
space is one-dimensional, and a single internal state is
associated with the ``bare dibaryon'' having the energy $E_D$.

The external Hamiltonian is
represented as a sum of three terms:
 \begin{equation}
 h_{NN}=h_{NN}^{0}+V_{\rm OPEP}+V_{\rm orth},
 \end{equation}
where $h_{NN}^{0}$ is the two-nucleon kinetic energy operator,
%(including also the Coulomb interaction for two protons)
$V_{\rm
OPEP}$ is the one-pion exchange potential which determines the peripheral
interaction of two nucleons, and $V_{\rm orth}$ is an orthogonalizing potential needed to exclude the fully symmetric six-quark component from the internal state wavefunction (see below).

Here we use the
same form and the same parameters of $V_{\rm OPEP}$ as in
Refs.~\cite{Yaf2019,Kukulin1}:
\begin{equation}
V_{\rm OPEP}=
-\frac{f_{\pi}^2}{m_{\pi}^2}({\bm\tau}_1\cdot {\bm\tau}_2)\frac{({\bm\sigma}_1\cdot
{\bf q})({\bm \sigma}_2\cdot {\bf q})}{q^2+m_{\pi}^2}\left(\frac{\Lambda_{\pi
NN}^2-m_{\pi}^2}{\Lambda_{\pi NN}^2+q^2}\right)^2. \label{Vope}
\end{equation}
In the calculations below, we employ the soft cutoff value $\Lam_{\pi NN}=0.65$
GeV/$c$~\cite{Yaf2019,Kukulin1}. This value is consistent with the microscopic quark-model~\cite{Koepf,Gari} and the lattice-QCD~\cite{Liu} calculations, as well as with the values commonly used in calculations of pion production (see~\cite{NPA2016} and references therein).

The potential $V_{\rm orth}$ has the separable form
\begin{equation}
V_{\rm
orth}=\lam_0|\phi_0\rangle\langle \phi_0|. \label{vort}
\end{equation}
This term corresponds to an effective repulsion and reflects the
six-quark symmetry requirements. In particular, the total
microscopic six-quark wavefunctions for the $S$-wave $NN$ interaction
include two different (mutually orthogonal) six-quark Young
schemes: $|s^6[6]\rangle$ and $|s^4p^2[42]\rangle$ in the quark
shell model language. It was shown in a number of works (see,
e.g., \cite{e1}) that while the fully symmetric component
$|s^6[6]\rangle$ describes a bag-like structure with the
hidden-color components, the mixed-symmetry component
$|s^4p^2[42]\rangle$ is projected mainly onto the $NN$ channel. So
that, to exclude the non-clustered bag-like component
$|s^6[6]\rangle$ from the $NN$ relative-motion wavefunctions, one
employs the orthogonalising projection operator
$\lam_0|\phi_0\rangle\langle \phi_0|$ with a large constant
$\lam_0$ \cite{e2}. This orthogonalisation of the physical $NN$
channel to the fully space-symmetric six-quark channel immediately
leads to the nodal $NN$ radial wavefunctions in the lower partial
waves with a stable radial node against change of the collision
energy \cite{e3}. It was argued in Ref.~\cite{Kukulin2} that the
coupling constant $\lam_0$ should be finite for the $S$-wave $NN$
interaction to take into account the strong coupling between $NN$
and $NN^*$(1440) channels near the Roper resonance $N^*$(1440)
excitation threshold. On the other hand, the term (\ref{vort})
leads to an appearance of some excited states in the total
six-quark system.

After excluding the internal channel by the standard projection technique (see details in
Refs.~\cite{Yaf2019,Kukulin1}), one gets an effective Hamiltonian in the $NN$ channel
with the main attraction given by the energy-dependent
pole-like interaction $V_D(E)$:
\begin{equation}
H_{\rm eff}(E)=h_{NN} + V_D(E), \quad V_D(E)=\frac{\lam^2}{E-E_D}|\phi\rangle \langle
\phi|, \label{Heff}
\end{equation}
where $\lam$ is a strength of coupling between the external and
internal channels and $|\phi\rangle$ is a transition form factor.
According to the symmetry requirements, the radial parts of the
form factors $|\phi_0\rangle$ and $|\phi\rangle$ are taken as the
lowest and the first excited harmonic oscillator wavefunctions for
the given orbital angular momentum $L$ and the same effective range $r_0$:
\begin{equation}
\label{phi0}
\phi_0^L(k)=\sqrt{\frac{2r_0(kr_0)^{2L+2}}{\Gamma(L+\frac{3}{2})}}e^{-\frac{1}{2}(kr_0)^2},
\end{equation}
\begin{equation}
\label{phi1}
\phi^L(k)=\sqrt{\frac{2r_0(kr_0)^{2L+2}}{\Gamma(L+\frac{5}{2})}}
\left[L+\frac{3}{2}-(kr_0)^2\right]e^{-\frac{1}{2}(kr_0)^2},
\end{equation}
where $\Gamma(L+\frac{3}{2})$ and $\Gamma(L+\frac{5}{2})$ are the
Gamma-functions.
%(see details in Ref.~\cite{Kukulin2}).
In principle, these form factors can be found from the microscopic
calculations within the six-quark models as it was done, e.g., in Refs.~\cite{JPhys2001,KuInt}.
However the full and consistent six-quark calculations are still beyond our capabilities, so that,
we have to consider the scale parameters $r_0$ of the form factors and also the coupling constants $\lam$ and
$\lam_0$ as adjustable parameters fitted to the data. Nevertheless, we tried to keep these parameter values (see Tabs.~\ref{Tab1} and \ref{Tab3}) as near as possible to the six-quark model and physical estimations\footnote{See, e.g., Ref.~\cite{KuInt} where a direct comparison of the potential parameters found from the fit of the phase shifts with their microscopic estimations
 was given for the lowest $NN$ partial channels $^3S_1$--$^3D_1$ and $^1S_0$ within the initial version of the dibaryon model.}.

\subsection{Treatment of inelastic processes}
In general, the internal six-quark state is able to decay into all possible
inelastic (other than $NN$) channels, including
meson and isobar production. In our model, this internal
state is considered as a ``bare'' dibaryon resonance state. Being
coupled to the $NN$ channel, this initial state gets to be
additionally ``dressed'' by the $NN$ loops in some analogy with
the field theory.

To take into account the possible inelastic decay channels,
we introduce the width $\Gamma_{\rm inel}$ for the
internal state energy $E_D$. We suppose here the energy dependence
$\Gamma_{\rm inel}(E)$ according to the phase space of the dominant decay
 channel and the relative orbital angular momenta of the final particles. This dibaryon
width has been introduced in our previous work
\cite{Yaf2019,Kukulin1} to allow for the treatment of both elastic and inelastic $NN$
scattering within the framework of the unified model.

Thus, the internal-state energy becomes complex-valued:
\begin{equation}
E_D=E_0-i\Gamma_{\rm inel}/2.
\end{equation}
Generally,
the width for the given decay channel can be written in the form:
   \begin{equation}
\Gamma_{\rm inel}(\sqrt{s})=\left\{
\begin{array}{lr}
0,& \sqrt{s}\leq E_{\rm thr};\\\displaystyle
\Gamma_0\frac{F(\sqrt{s})}{F(M_0)},&\sqrt{s}>E_{\rm thr}\\
\end{array}\label{gamd}
\right.,
\end{equation}
where $\sqrt{s}$ is the total invariant energy of the decaying
resonance, $M_0$ is the bare dibaryon mass related to the energy $E_0$ as $M_0=2\sqrt{m(E_0+m)}$ (with $m$ being the proton mass), $E_{\rm
thr}=2m+m_\pi$ is the threshold energy and $\Gamma_0$ defines
the decay width at the resonance position. The function $F(\sqrt{s})$ depends on the
type of the decay process.

For the $NN$ channels in question, the dominant decay process
close to the pion production threshold should be just the emission
of the neutral or charged pion, i.e., $D\to NN \pi$. The
explicit expression for the function $F$ for such a three-body
decay can be taken in the following form \cite{Yaf2019}:
\begin{eqnarray}
F_{NN\pi}(\sqrt{s})=
\frac{1}{s}\int_{2m}^{\sqrt{s}-m_{\pi}}dM_{NN} \nonumber\\
\frac{q^{2l_\pi+1}k^{2L_{NN}+1}}{(q^2+\Lam^2)^{l_\pi+1}(k^2+\Lam^2)^{L_{NN}+1}},
\label{fpinn}
\end{eqnarray}
where  $\displaystyle
q={\sqrt{(s-m^2_\pi-M^2_{NN})^2-4m_\pi^2M_{NN}^2}}\Big/{2\sqrt{s}}$
is the pion momentum in the total center-of-mass frame,
$\displaystyle k=\half\sqrt{M_{NN}^2-4m^2}$ is the momentum of
the nucleon in the center-of-mass frame of the final $NN$
subsystem with the invariant mass $M_{NN}$, and $\Lam$ is the
high-momentum cutoff parameter which prevents an unphysical rise
of the width $\Gamma_{\rm inel}$ at high energies. The orbital angular
momenta of the emitted pion $l_{\pi}$ and the nucleon in the $NN$ subsystem $L_{NN}$
may take different values in accordance with the total angular momentum and parity conservation. For the $^3P_0$ initial channel, we parameterize the dibaryon width according to the dominant near-threshold $Ss$ final state with $L_{NN}=l_{\pi}=0$, while for the $^3P_2$ channel we take $L_{NN}=0$ and $l_{\pi}=2$ which correspond to the final $Sd$ configuration.

It should be noted that for the $^3P_2$ and $^3P_1$ channels, the decay process $D\to d\pi$ with the
final deuteron is also important near the inelastic threshold. In
such a case, the function $F$ takes a simple form corresponding to
the two-body decay:
\begin{equation}
\label{F_dpi}
F_{d\pi}(\sqrt{s})=\frac{q^{2l_\pi+1}}{(q^2+\Lam_{d\pi}^2)^{l_\pi+1}},
\end{equation}
where $q$ is the pion momentum and $\Lam_{d\pi}$ is the high-momentum cutoff parameter. It can be shown that the
general three-body expression (\ref{fpinn}) can be reduced to the
two-body form similar to Eq.~(\ref{F_dpi}) in the case of small
relative momenta of two emitted nucleons due to the close
values of the deuteron mass $m_d$ and the two-nucleon threshold
energy $M_{NN}=2m$. Thus, one can use the three-body
parametrization of $\Gamma_{\rm inel}$ (\ref{gamd}) for all three isovector $P$-wave $NN$ channels
(with the function $F$ defined by Eq.~(\ref{fpinn})) as the
most common form of the decay width which effectively takes into account
the possible inelastic processes for these channels near the inelastic threshold. For the description of $NN$ scattering phase shifts at higher energies, the energy behaviour of the dibaryon decay width is much less important, so that, Eq.~(\ref{fpinn}) can be used to represent the width also above the threshold region.

The bare dibaryon mass $M_0$ and width $\Gamma_0$ are renormalised in the course of dressing by the $NN$ loops, i.e., when solving the scattering equations with the effective Hamiltonian (\ref{Heff}), thus resulting in the ``dressed'' dibaryon mass $M_{\rm th}$ and width $\Gamma_{\rm th}$ which can be compared to the experimental dibaryon position. The inelastic width at the resonance point $\Gamma_{\rm inel}(M_{\rm th})$ determines the coupling of the dibaryon resonance in question to the inelastic channels, the main one being the $NN \pi$ channel. Accordingly, the elastic decay width of the dibaryon can be found as $\Gamma(D\to NN)=\Gamma_{\rm th}-\Gamma_{\rm inel}(M_{\rm th})$ (see details in Ref.~\cite{Yaf2019}).

\section{Description of $\bm{pp}$ scattering in $\bm{P}$ waves}

\subsection{Partial phase shifts and inelasticities in the channels $\bm{^3P_0}$ and $\bm{^3P_2}$--$\bm{^3F_2}$}

Below we consider the results of calculations using the above
formalism for the $pp$ partial-wave channel $^3P_0$ and the coupled channels $^3P_2$--$^3F_2$.
%\footnote{Although we discuss here mainly the
%proton-proton scattering the Coulomb interaction is omitted for a
%sake of simplicity in the present work. It should be mentioned
%that in our model it doesn't play an important role in the
%intermediate energy region at which our study is focused.}
The parameters of the model used in the present calculations are listed in Tab.~\ref{Tab1}.
Here a single internal state for the coupled channels $^3P_2$ and $^3F_2$ is introduced (see the details below).
The mass $m_{\pi}$ has been taken to be equal to the neutral pion mass. The main
parameters of the $P$-wave dibaryon resonances in the effective
Hamiltonian (\ref{Heff}), i.e., the mass $M_0$ and width
$\Gamma_0$ of the bare resonance, have been fitted to reproduce
the partial phase shifts and inelasticity parameters in a broad energy
interval up to the resonance positions for the given $NN$
channels.
\begin{table}[h]
\caption{Parameters of the dibaryon model potential  for the $NN$
partial-wave channels $^3P_0$ and $^3P_2$--$^3F_2$.\label{Tab1}}
\begin{tabular}{cccccc}
\hline
&$\lam_0$&$r_0$&$\lam$&$M_0$&$\Gamma_0$ \\
& MeV&fm&MeV&MeV&MeV\\
\hline
$^3P_0$& 450 & 0.425 & 35 & 2200 & 92  \\
\hline
$^3P_2$& 0 & 0.7 & 63 &\multirow{2}*{2205} &\multirow{2}*{100} \\
$^3F_2$ & 105& 0.45& 1.5 &  & \\
\hline
\end{tabular}
\end{table}

With these model parameters, one gets the ``dressed'' resonance states in both
configurations with the masses $M_{\rm th}$ and widths $\Gamma_{\rm
th}$ shown in Tab.~\ref{Tab2}. We emphasise that these values occur to be fully
consistent with the experimental ones found in
Ref.~\cite{Komarov16}.\footnote{The $F$-wave admixture to the $^3P_2$ state was not considered in the experimental work~\cite{Komarov16}.} So, these two basic parameters of the dibaryon resonances in our model can be fixed by the experimental data.

\begin{table}[h]
\caption{Comparison of the experimental values for the resonance
masses and decay widths with those found in the dibaryon model
for the $NN$ partial-wave channels $^3P_0$ and $^3P_2$--$^3F_2$.
\label{Tab2}}
\begin{tabular}{ccccc}
\hline
& $M_{\rm th}$& $\Gamma_{\rm th}$&$M_{\rm exp}$&$\Gamma_{\rm exp}$\\
& MeV&MeV&MeV&MeV\\
\hline
$^3P_0$ & 2200 & 99 & 2201(5) & 91(12) \\
$^3P_2$--${^3F_2}$ & 2221 & 168 & 2197(8) &130(21)\\
\hline
\end{tabular}
\end{table}

Using Eqs.~(\ref{gamd}), (\ref{fpinn}) and the values of the dibaryon mass and width given in Tab.~\ref{Tab2}, one can estimate the branching ratio for the dibaryon decay into the $NN$ channel: ${\rm BR}(D\to NN)=(\Gamma_{\rm th}-\Gamma_{\rm inel}(M_{\rm th}))/\Gamma_{\rm th}$~\cite{Yaf2019}. Thus, we obtain this branching ratio to be about 7\% and 13\% for the $^3P_0$ and $^3P_2$--$^3F_2$ dibaryons, respectively.
These values are consistent with the value of 10\% obtained for the $P$-wave dibaryons from the PWA of $pp$ scattering~\cite{Strak3P1,Strak91}.
So, these dibaryons are highly inelastic as well as other known dibaryons above the pion production threshold~\cite{hcl}.

\subsubsection{The channel $^3P_0$}

\begin{figure}[h]
\centering \epsfig{file=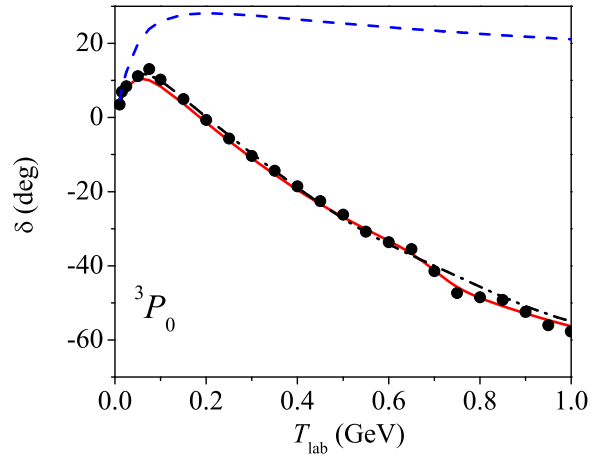,width=\columnwidth}
\caption{\label{fig3P0} \small (Color online) Partial phase shifts
for the $pp$ channel $^3P_0$ found within the dibaryon model
(solid curve) in comparison with the single-energy SAID
PWA~\cite{SAID} (filled circles), the SAID SM16 solution~\cite{SAID}
(dash-dotted curve) and results for the pure OPEP
(dashed curve).}
\end{figure}

The partial phase shifts for the $pp$ channel $^3P_0$ are
displayed in Fig.~\ref{fig3P0} in comparison with the
PWA results of the SAID group \cite{SAID}.

As is clearly seen from
the Figure, we have achieved almost perfect agreement with the results of PWA in a very broad
energy range from zero energy to about 1 GeV. To make the result more evident, we have also shown on
Fig.~\ref{fig3P0} the phase shifts corresponding to the pure OPEP
(dashed curve). This comparison makes clear the fact that the
dibaryon formation in our model provides the dominant part of the $NN$ interaction in
the whole energy region considered.

\begin{figure}[h]
\centering \epsfig{file=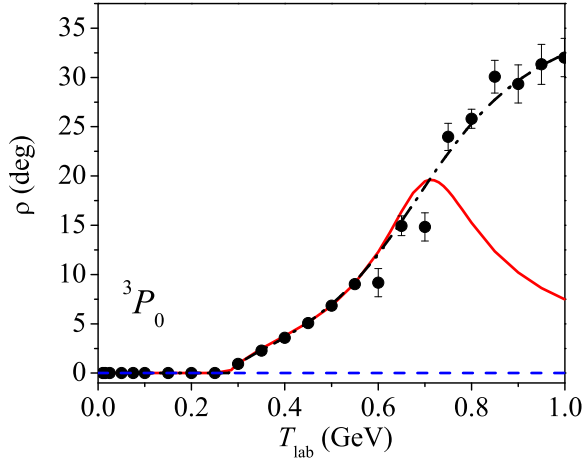,width=\columnwidth}
\caption{\label{fig3P0_inel} \small (Color online) Inelasticity
parameters for the $pp$ channel $^3P_0$ found within the dibaryon
model (solid curve) in comparison with the single-energy SAID
PWA~\cite{SAID} (filled circles), SAID SM16 solution~\cite{SAID} (dash-dotted
curve) and results for the pure OPEP (dashed curve).}
\end{figure}

To take an effective account of inelastic processes, we have used here
Eqs.~(\ref{gamd}) and (\ref{fpinn}) for the internal state decay
width corresponding to the single-pion production. The resulted inelasticity parameters are shown in Fig.~\ref{fig3P0_inel}.
Again, a very good agreement with the PWA results is seen up to
the energies near the resonance position. Further, the total
inelasticity continues to increase signaling about other inelastic
processes (such as double-pion production, etc.) while the
inelasticity predicted by the dibaryon model decreases.

We emphasize here that the excellent description of both
elastic phase shifts and inelasticities shown in
Figs.~\ref{fig3P0} and~\ref{fig3P0_inel} has been attained using the
same pole parameters, i.e., within a unified model.

\subsubsection{The coupled channels $^3P_2$--$^3F_2$}
The preliminary study of the $NN$ scattering phase shifts and
inelasticity parameters in the $^3P_2$
channel within the dibaryon model has already been published in Ref.~\cite{Yaf2019}.
Here they are explored in a more detail with inclusion of coupling with the $^3F_2$ channel and focusing
on the description of inelastic processes. Due to account of the $PF$ coupling, the dibaryon part of the interaction in Eq.~(\ref{Heff})
should have the following form:
\begin{equation}
V_{D}(E)=\frac{1}{E-E_D}|\Phi\rangle\langle \Phi|,\quad |\Phi\rangle=
\left(\begin{array}{c}
\lam_P|\phi_P\rangle\\
\lam_F|\phi_F\rangle\\
\end{array}\right),\label{vd2}
\end{equation}
where $|\Phi\rangle$ contains two form factors corresponding to coupling of the internal state with two partial $NN$ channels in question.
Eq.~(\ref{vd2}) results in the following matrix form of the interaction:
\begin{equation}
V_D(E)=\frac1{E-E_D}
\left(
\begin{array}{cc}
\lam_P^2|\phi_P\rangle\langle \phi_P|& \lam_{PF}^2|\phi_P\rangle\langle \phi_F|\\
\lam_{PF}^2|\phi_F\rangle\langle \phi_P|&\lam_F^2|\phi_F\rangle\langle \phi_F|\\
\end{array}
\right),
\end{equation}
where $\lam_{PF}=\sqrt{\lam_P\lam_F}$ and the non-diagonal parts give impact to the tensor component of the total interaction.

In Eq.~(\ref{vd2}), the partial form factors are taken again as the harmonic oscillator functions
 corresponding to the orbital angular momentum $L$ and the
effective range $r_0$. However, in contrast to the $^3P_0$ channel, the $^3P_2$ partial phase shifts are positive and do not show any
 repulsion until the energy of
about 1 GeV. Hence, the orthogonalizing potential is needed in the
$F$-wave channel only. So that, the form factor $|\phi_P\rangle$ represents the lowest harmonic oscillator wavefunction given by Eq.~(\ref{phi0}) with $L=1$,
 while the form factors $|\phi_0\rangle$ and $|\phi_F\rangle$ represent the lowest and the first excited harmonic oscillator functions defined in Eqs.~(\ref{phi0}) and (\ref{phi0}), respectively, with $L=3$. The model parameters for these coupled channels are collected in Tab.~\ref{Tab1}.

\begin{figure}[h]
\centering \epsfig{file=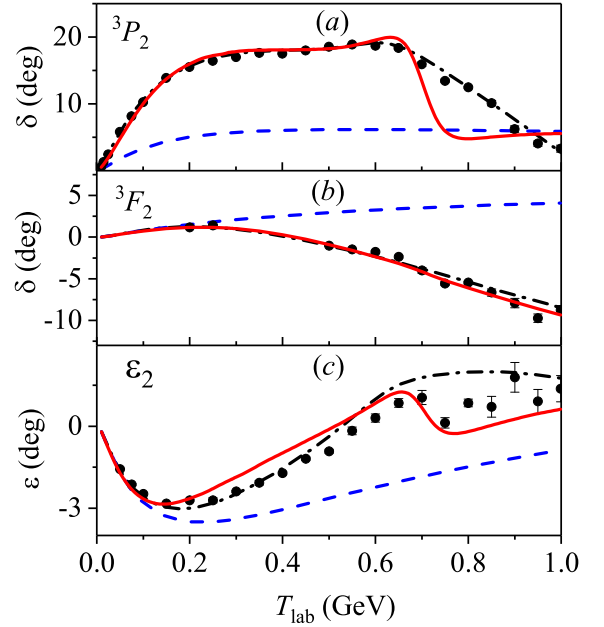,width=0.9\columnwidth}
\caption{\label{phase_3P2} \small (Color online) Partial phase shifts for the coupled $pp$ channels $^3P_2$ (a) and $^3F_2$ (b) and the mixing angle $\varepsilon_2$ (c). The notations are the same as in Fig.~\ref{fig3P0}.}
\end{figure}

The $^3P_2$ and $^3F_2$ $pp$ scattering phase shifts and the mixing angle $\varepsilon_2$ are shown in
Fig.~\ref{phase_3P2} in comparison with the SAID PWA~\cite{SAID}
and the pure OPEP contribution.
It is seen that the phase shift $^3F_2$ is represented quite
well up to 1 GeV. However, the good
description of the partial phase shifts in the $^3P_2$ channel is obtained  until ca. 700
MeV (i.e., up to the resonance position) only. The coupling between the $P$ and $F$ waves is small here which results in a rather small mixing angle. As is seen from Fig.~\ref{phase_3P2}$c$, this mixing angle is defined essentially by the tensor part of the OPEP.
 At the same time, the tensor part of the dibaryon potential gives the required correction near the resonance position. However, its contribution turns out to be slightly overestimated in the intermediate energy region.

\begin{figure}[h]
\centering \epsfig{file=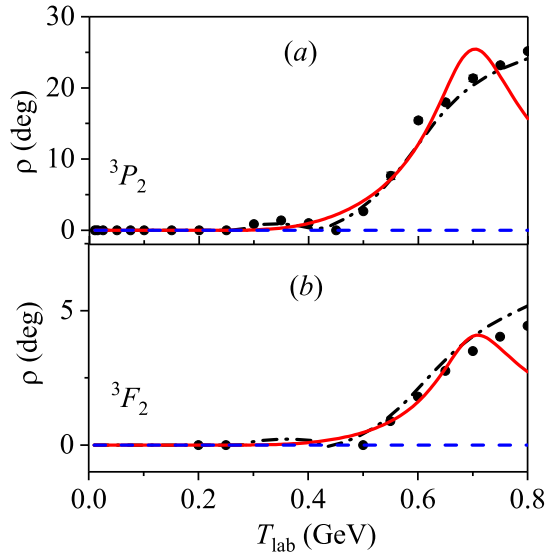,width=0.9\columnwidth}
\caption{\label{inel_3P2} \small (Color online) Inelasticity parameters for the coupled $pp$ channels $^3P_2$ (a) and  $^3F_2$ (b). The notations are the same as in Fig.~\ref{fig3P0_inel}.}
\end{figure}

The respective inelasticity parameters found within the dibaryon model
are shown in Fig.~\ref{inel_3P2}. Although some overestimation of
the inelasticity is seen in $^3P_2$ channel in the intermediate energy region, one
can get a reasonable average description of the inelastic processes in both $P$ and $F$ partial waves.

To summarize the comparison with the PWA data, it should be emphasized that a rather simple form of the interaction with a single internal state allows us to reproduce quite well {\em all five scattering parameters} (i.e., the phase shifts, the inelasticities and mixing angles)  corresponding to the coupled $NN$ channels
$^3P_2$--$^3F_2$ and simultaneously gives the resonance position (see Tab.~\ref{Tab2}) which is very close to the experimental one \cite{Komarov16}. At the same time, some deviations from the PWA results are seen which should stimulate some further improvements of the model.

In particular, small discrepancies between our calculations and the PWA data
in the near-threshold region might be explained by the more involved structure of the dibaryon decay width in the $^3P_2$
channel compared to the $^3P_0$ one, where the final $Ss$ configuration dominates the near-threshold pion production. For the $^3P_2$
initial channel, the $Sd$, $Ds$ and $Pp$ final configurations can give comparable contributions, while we take into account the $Sd$ configuration only (which is appropriate for the $d\pi^+$ final state). Thus, the correct
description of the total inelasticities in the $^3P_2$--$^3F_2$ partial channels requires a detailed dynamical model for the dibaryon decay in various final states instead of our simple width parametrization. Such a dynamical description of all dibaryon decay channels is beyond the scope of the present work.

\subsection{Partial phase shifts and inelasticities in the channel $\bm{^3P_1}$}

For consistency of our study of the $P$-wave $pp$ scattering within the dibaryon model, we should examine the model in the $^3P_1$ partial channel as well. Though the $^3P_1$ dibaryon resonance has not been detected in experiments to date, different
estimations from the PWA of $pp$ scattering can be found in the literature. In particular, in Ref.~\cite{Strak3P1}, the resonance state with the mass $M({}^3P_1)=2179$ MeV and the width $\Gamma({}^3P_1)=86$ MeV was found from the PWA of $pp$ scattering.
At the same time, the hypothetical $^3P_1$ dibaryon state with the mass $M({}^3P_1) \simeq 2.07$ GeV can be considered as a member
 of the Regge trajectory for isovector dibaryons \cite{MacGregor}. The resonance with the same mass was also mentioned in Ref.~\cite{Ferreira}.

Below we show the results of the calculations of the $^3P_1$ $pp$ phase shifts within the dibaryon model for one of the possible parameter sets
(see Tab.~\ref{Tab3}).
\begin{table}[h]
\caption{Parameters of the dibaryon model potential for the $NN$
partial-wave channel $^3P_1$.\label{Tab3}}
\begin{tabular}{cccccc}
\hline
&$\lam_0$&$r_0$&$\lam$&$M_0$&$\Gamma_0$ \\
& MeV&fm&MeV&MeV&MeV\\
\hline
$^3P_1$& 270 & 0.425 & 20 & 2230 & 50   \\
\hline
\end{tabular}
\end{table}
Here, the effective range $r_0$ of the dibaryon part of the interaction is the same as for the channel $^3P_0$ (see Tab.~\ref{Tab1}).

\begin{figure}[h]
\centering \epsfig{file=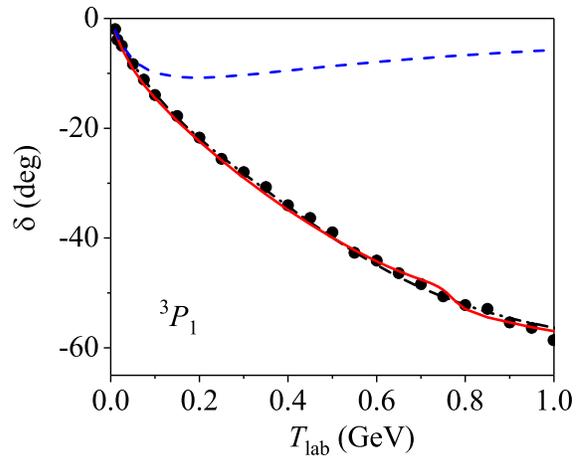,width=\columnwidth}
\caption{\label{phase_3P1} \small (Color online) Partial phase shifts for the $pp$ channel $^3P_1$. The notations are the same as in Fig.~\ref{fig3P0}.}
\end{figure}

\begin{figure}[h]
\centering \epsfig{file=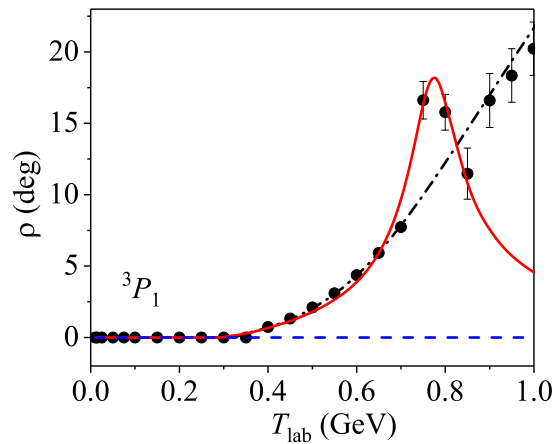,width=\columnwidth}
\caption{\label{inel_3P1} \small (Color online) Inelasticity parameters for the $pp$ channel
$^3P_1$. The notations are the same as in Fig.~\ref{fig3P0_inel}.}
\end{figure}

The partial phase shifts and inelasticity parameters for the channel $^3P_1$ are shown in Figs.~\ref{phase_3P1} and \ref{inel_3P1},
respectively, in comparison with the SAID PWA data. Similarly to the $^3P_0$ case, one obtains the good description of these
quantities up to laboratory energies of about 1 GeV.
Moreover, with the parameters chosen, the model allows us to reproduce the peak that appears in the SAID single-energy data near $T_{\rm lab}=0.75$ GeV. It should be noted however, that the energy-dependent SAID solution does not reveal such a peak.

Finally, the suggested dibaryon potential results in the ``dressed'' dibaryon resonance with the mass
 $M_{\rm th}({}^3P_1)=2230$ MeV and the width $\Gamma_{\rm th}({}^3P_1)=52.5$ MeV. Thus, the total width of the $^3P_1$ dibaryon in our model turns out to be narrower than those for other $P$-wave dibaryons. For the branching ratio of the $^3P_1$ dibaryon decay into the $NN$ channel we obtain the value of about 5\% which is somewhat smaller than for the other $P$-wave dibaryons as well.
We should note however, that the parameters of the model potential and the resulted dibaryon resonance which allow us to describe reasonably $pp$ scattering in the $^3P_1$ channel are not unique. So, additional support from the experimental side is needed to draw unambiguous conclusions about the existence of the $^3P_1$ dibaryon.

%\section{Pion production in $\bm{pp}$ collisions via the intermediate dibaryon resonances}
\section{Description of near-threshold neutral pion production}

In Sect. III A above, we have shown within the dibaryon model that the resonance ${}^3P_0(2200)$ found clearly in the experiment~\cite{Komarov16} determines the $pp$ inelasticity in the ${}^3P_0$ channel from the inelastic threshold to about $T_{\rm lab}=700$ MeV. In turn, the near-threshold inelasticity here is mainly due to the neutral pion production, since
the process $pp\to pp\pi^0$ in the near-threshold region is by far dominated by the $^3P_0$ initial state of the $pp$ pair which is coupled to the $^1S_0$ $pp$ final state with an $s$-wave emitted pion (the so-called
$Ss$ partial amplitude) \cite{WASA_M}.\footnote{For the initial $pp$ pair in the $^3P_0$ state, there is also a small admixture of the $^3P_1p$ (or $Pp$) final configuration which is rising with the collision energy \cite{WASA_M}.} On the other hand, the total inelastic $pp$ cross section in the $^3P_0$ channel consists of two parts corresponding to the above neutral pion production
process and also the charged pion
production reaction $pp\to pn\pi^+$. Due to the different dynamics of these processes, including the different final state interaction of the $pp$ and $np$ pairs with the small relative momenta, it seems
nontrivial to extract these amplitudes separately. However, the
threshold of the charged pion production is shifted by about 10 MeV (in the lab. energy)
from that of the neutral pion production. Thus, one has a unique situation in the small energy
interval just above the neutral pion production
threshold, where the total cross section of the $pp\to
pp\pi^0$ reaction should closely coincide with the total inelastic
cross section in the $pp$ partial channel $^3P_0$. Thus, the dibaryon resonance which governs $pp$ scattering in the $^3P_0$ channel, should determine also near-threshold $\pi^0$ production in $pp$ collisions.

\begin{figure}[ht]
\centering \epsfig{file=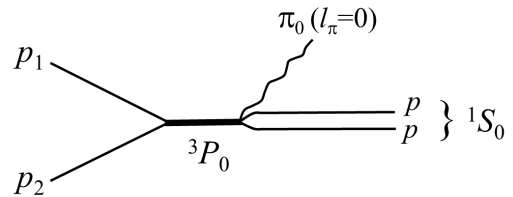,width=0.8\columnwidth}
\caption{\label{fig3} The dibaryon-induced  mechanism for
near-threshold $\pi^0$ production in the $^3P_0$ $pp$ channel
which leads to the $^1S_0$ $pp$ final state. The microscopic
interpretation of the production mechanism in terms of the
$4q$--$2q$ six-quark model is presented in Sect. V.}
\end{figure}

The proposed dibaryon mechanism for near-threshold $\pi^0$ production is depicted in Fig.~\ref{fig3}.

The fundamental difference between the traditional mechanism for
the $\pi^0$ production displayed in Fig.~\ref{fig1} and the
dibaryon-induced mechanism for this process shown in
Fig.~\ref{fig3} is as follows. The dominating $NN$ channel leading
to the $\pi^0$ near-threshold production is $^3P_0$ which is odd
with respect to the two-nucleon permutation. So that, for the
single-nucleon contributions to the $\pi^0$ production from the
$^3P_0$ initial channel of the type shown in Fig.~\ref{fig1} a
strong mutual cancellation between the production on the first and
second nucleons takes place, while the dibaryon mechanism shown
in Fig.~\ref{fig3} has no such mutual single-particle
cancellation.

%Below we present the
%results of the calculation for the near-threshold $\pi^0$ production
%based on the mechanism depicted in Fig.~\ref{fig3}.

\begin{figure}[h]
\centering \epsfig{file=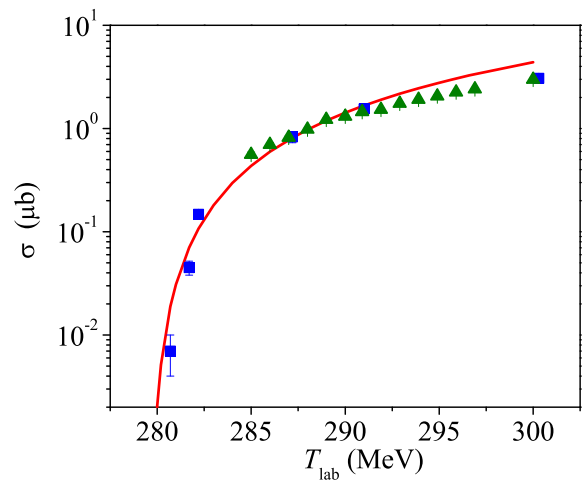,width=\columnwidth}
\caption{\label{Meyer} \small (Color online)  The total inelastic
$pp$ cross section in the $^3P_0$ channel found within the
dibaryon model (solid curve) in comparison with experimental data on the total
$pp\to pp\pi^0$ cross section from
Refs.~\cite{Meyer} (triangles) and \cite{Bondar} (squares).}
\end{figure}

In Fig.~\ref{Meyer} we compare the total inelastic cross
section found for the $^3P_0$ channel within our model with the existing
experimental data for the near-threshold $\pi^0$ production \cite{Meyer,Bondar}.

Very good agreement between our calculations and experimental data is seen in the energy interval
$T_p=$280--290 MeV just above the neutral pion production
threshold. It should be emphasized that although some small
discrepancies with the data are visible in some experimental points,
the general behavior of the cross section as a function of the collision energy
is reproduced quite well. On the one hand, it is not surprising to have a good description of the near-threshold $\pi^0$ production provided the total inelasticity of $pp$ scattering is fitted in the $^3P_0$ channel. On the other hand, it was not evident at all that a simple model with a single dibaryon pole located rather far from the $\pi NN$ threshold could reproduce the $pp$ inelasticity in a broad energy range, including a very good description of the near-threshold region. At higher energies, above the charged-pion production threshold, the total
inelastic $pp$ cross section overestimates the experimental
data on the neutral pion production, due to the rising contribution of the process $pp\to pn \pi^+$.

It has been demonstrated already in Ref.~\cite{Meyer} that the qualitative behavior of the near-threshold neutral pion
production cross section can be understood in terms of the three-body phase space multiplied by a FSI factor.
On the other hand, in the current version of the dibaryon model, an account of inelastic processes is
done by incorporating the decay width $\Gamma_{\rm inel}$. Thus, the
success of the model in the description of the near-threshold pion production cross section is
due to the proper behaviour of the function (\ref{fpinn}) at small
pion momenta $q$. For the $Ss$ channel, this function indeed behaves as a three-body phase space with a cutoff factor, which effectively takes into account the FSI between the final protons. However, our results are more general since this width parametrization allows us to reproduce the $pp$ inelasticity not only near the pion production threshold but in a wide energy interval up to the dibaryon resonance position.

Thus, within the model suggested, we have a good description
of $pp$ elastic scattering in the $^3P_0$ partial channel
from zero energy up to about 1 GeV and the correct near-threshold behaviour of the neutral pion production cross section as
well. We can conclude therefore that $pp$ scattering in this channel is determined almost completely by the $^3P_0$ dibaryon resonance with a mass of about $2200$ MeV.

The $^3P_2$ $pp$ channel also plays an important role in the neutral pion production process~\cite{Komarov16}. However, the respective amplitude is small in the near-threshold region due to the angular momentum barrier. At the same time, at energies where it becomes larger, our theoretical total inelastic amplitude includes effectively not
only the $pp\pi^0$ channel but also $pn\pi^+$ and
$d\pi^+$ ones. Thus, we cannot extract the contribution of the $^3P_2$ dibaryon to the particular inelastic processes within the current version of the model. However, we have got a reasonable description of the elastic phase shift and the total inelasticity in the $^3P_2$ channel as well. Besides that, our study of the reaction $pp\to d\pi^+$ by inclusion the dibaryon resonances into the leading partial-wave amplitudes has
shown the dominance of the $^3P_2$ dibaryon in the respective partial cross section (see Ref.~\cite{PRD2016} and Fig.~\ref{fig-3p2d} of the present paper). These results allow us to suppose the leading role of the dibaryon mechanism in $pp$ elastic scattering and pion production processes in the $^3P_2$ channel from threshold up to the resonance position.

Finally, the $^3P_1$ dibaryon, if it exists, should determine the near-threshold charged pion production processes in $pp$ collisions with the isoscalar $np$ (or the deuteron) final state, i.e., $pp \to \{pn\}_{I=0} \pi^+$ and $pp \to d \pi^+$. However, it is non-trivial to separate these processes with the very near thresholds having just the total inelastic cross section in the $^3P_1$ channel. So, we restrict ourselves to the reasonable description of the scattering phase shift and the total inelasticity in this channel as well.

\section{Microscopic structure of $\bm{P}$-wave dibaryon resonances}

In the previous section, we demonstrated that the $^3P_0$ and $^3P_2$ isovector dibaryon resonances discovered recently by the
ANKE-COSY Collaboration \cite{Komarov16} govern both $NN$ elastic and inelastic scattering in
the respective partial-wave channels from zero energy until at least 700 MeV. We have also shown that the near-threshold
production of neutral pions is determined mainly by the $^3P_0$ dibaryon resonance.

\begin{figure}[h]
\centering \epsfig{file=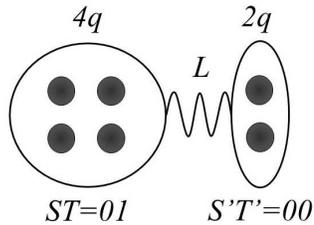,width=0.5\columnwidth}
\caption{\label{struct} \small The tetraquark-diquark model for the dibaryon state with the colour
string connecting two quark clusters.}
\end{figure}

On the other hand, we discussed in Ref. \cite{NPA2016}
some possible realisation of the Nijmegen--ITEP quark-cluster model \cite{Nijm,ITEP} for the
well-known isovector dibaryon resonance trajectory $^1D_2$, $^3F_3$, $^1G_4$, etc. The feasible realisation
of the general $4q$--$2q$ model for the straight-line isovector dibaryon trajectory is the six-quark
structure shown in Fig.~\ref{struct}. Here $S$ and $T$ are the spin and isospin of the tetraquark $4q$ while the respective
values for the diquark $2q$ are denoted as $S'$ and $T'$.

In the six-quark model proposed here, i.e., with $ST=01$ and $S'T'=00$,
the total angular momentum is equal to the orbital angular momentum, i.e., $J=L$, and the total isospin $T_{\rm tot}=1$. In this way, we can describe the whole isovector trajectory $^1S_0$, $^3P_1$, $^1D_2$, $^3F_3$, $^1G_4$, etc., considered in Ref.~\cite{MacGregor} as rotational excitations in the $NN$ system, by a rotating color string between the $4q$ and $2q$ clusters with the orbital angular momenta $L=0,1,2,\ldots$, respectively.\footnote{Note that in Ref. \cite{NPA2016} the quantum numbers $ST$ and $S'T'$ were selected in a different way which did not allow for a description of two lowest members of the isovector dibaryon trajectory, i.e., $^1S_0$ and $^3P_1$. The $^3P_0$ and $^3P_2$ resonances also did not fit into that classification, so the three-diquark structure was suggested for them in Ref. \cite{PRD2016}.} Most of these isovector dibaryons have been detected to date, except for the $^3P_1$ resonance which should have the mass of about 2.07 GeV to fit into the above trajectory~\cite{MacGregor}. On the other hand, in the present study we have obtained the $^3P_1$ dibaryon resonance with the mass $M_{\rm th}({^3P_1})=2230$ MeV, i.e., close to the $^3P_0$ and $^3P_2$ resonances found experimentally. So, one can suppose that the $^3P_1$ resonance is shifted somehow by the spin-orbital forces which mix three $P$-wave states with different $J$, or, alternatively, that there are two $^3P_1$ dibaryons separated by about 150 MeV. However, since no one of these resonances has been detected to date, we restrict the following consideration to the $^3P_0$ and $^3P_2$ states.

Then the important question arises, i.e., could the microscopic structure within the framework of the above $4q$--$2q$ model be relevant also for the isovector dibaryon resonances $^3P_0$ and $^3P_2$? In fact, one can suggest
for the $^3P_0$ and $^3P_2$ states a similar $4q$--$2q$ microscopic structure but now with $ST=01$, $S'T'=11$ and $L=1$. This corresponds to the plausible quantum numbers of the $6q$ state, i.e., $S_{\rm tot}=1$,
$T_{\rm tot}=1$ and $J=L\pm1=0,2$. In such a case, two $P$-wave dibaryons $^3P_0$ and $^3P_2$ should have
just the $N$--$\Delta$ hadronic structure with the relative $P$ wave between the nucleon and the
$\Delta$-isobar. So, we have here an analogy with the structure of the well-known
$^1D_2(2150)$ dibaryon lying very near to the $N$--$\Delta$ threshold with a relative $S$-wave between $N$ and
$\Delta$.

Assuming further the above $4q$--$2q$ structure for the $^3P_0$ and $^3P_2$ dibaryons (with $ST=01$, $S'T'=11$ and $L=1$), one may explain
the $\pi^0$ emission from these dibaryons leading to a final $^1S_0$ diproton as a two-step process:
\begin{equation}
pp\to D({^3}P_{J=0,2})\to \{pp\}_{s}+\pi^{0}|_{l_{\pi}=0,2}.
\end{equation}
This process goes
through a conventional quark spin-isospin flip inside the axial-vector diquark: $S'T'=11\to 00$, with an accompanying deexcitation of the
colour string $L=1\to 0$ to compensate the parity change in the pion emission with $l_{\pi}=0$ or 2.

Thus, the suggested six-quark picture gives at least a qualitative explanation for the pion
emission induced by the intermediate $^3P_0$ and $^3P_2$ dibaryons.

\section{Conclusion}
In the present work, we studied the impact of the recently discovered \cite{Komarov16} $P$-wave dibaryon resonances with a mass of about 2200 MeV
on $pp$ elastic scattering and pion production in $pp$ collisions. Within the dibaryon-induced model for $NN$ interaction, we have demonstrated that these $P$-wave dibaryons give a quite reasonable description of the $pp$ scattering phase shifts and inelasticities in the $^3P_0$ and $^3P_2$--$^3F_2$ partial channels from zero energy to at least 700 MeV (lab.). Furthermore, we have shown that the $^3P_0$ dibaryon resonance determines almost completely the neutral pion production process $pp\to\{pp\}_{s}\pi^0$ in the near-threshold region. It should be stressed that the respective cross section does not find a satisfactory explanation within the conventional approaches. So, within the dibaryon model proposed, the amplitude of near-threshold $\pi^0$ production in $pp$
collisions is reproduced consistently with the $pp$ elastic scattering
amplitudes in the above partial-wave channels. The $^3P_2$ dibaryon resonance also gives an important contribution
to both neutral and charged pion production in a broad energy range, though its contribution is suppressed near threshold by the angular momentum barrier.

We have also achieved a very good description of the $pp$ scattering phase shift and inelasticity in the $^3P_1$ partial channel at $T_{\rm lab}=0$--700 MeV, using the dibaryon with a mass $M({}^3P_1) = 2230$ MeV, i.e., rather close to that of the $^3P_0$ and $^3P_2$ resonances found experimentally \cite{Komarov16}. If this dibaryon exists, it should determine the charged pion production processes $pp\to d\pi^+$ and $pp\to \{pn\}_{I=0}\pi^+$ near threshold. In fact, it seems natural to have three almost degenerate $P$-wave dibaryons with $J=0,1,2$ coupled with the spin-orbit force. It should be borne in mind however, that the $^3P_1$ dibaryon has not been detected to date, though some indications of its existence can be found in the literature. So, an additional experimental confirmation of these results is needed.

The results of the present work should be considered jointly with our previous conclusions \cite{PRD2016} about dominance of the $^3P_2$ dibaryon resonance in the charged pion production process $pp\to d\pi^+$ in the $^3P_2d$ partial wave and the crucial role of the dibaryon in the proper description of polarization observables in this process. A good description of both neutral and charged pion production as well as $NN$ elastic scattering within the unified dibaryon model gives a strong argument in favour of the decisive role of dibaryon resonances in $NN$ collisions
at intermediate energies in general.

\label{concl} {\bf Acknowledgments.} The work has been partially
supported by RFBR, grants Nos. 19-02-00011, 19-02-00014. M.N.P.
also appreciates support from the Foundation for the Advancement
of Theoretical Physics and Mathematics ``BASIS''.

\end{document}